\documentclass[table]{ccjnl}
\usepackage{lipsum,amsmath}
\usepackage{cuted}
\graphicspath{{figures/}}

\title{Passive Beamforming Design for Intelligent Reflecting Surface Assisted MIMO Systems}
\author{Chenghao Feng\inst{1,2}, Wenqian Shen\inst{1,*}, Xinyu Gao\inst{3}, Jianping An\inst{1}\corinfo{shenwq@bit.edu.cn}
}
\receiveddate{Aug. 06, 2020}
\reviseddate{Oct. 28, 2020 and Dec. 08, 2020}
\Editor{Caijun Zhong}

\address[1]{School of Information and Electronics, Beijing Institute of Technology, Beijing 100081, China}
\address[2]{Shaanxi Key Laboratory of Information Communication Network and Security, Xian University of Posts and Telecommunications, Xian, Shaanxi 710121, China}
\address[3]{Huawei Technologies Co. Ltd., Beijing 100085, China}

\begin{document}

\maketitle

\begin{abstract}
Intelligent reflecting surfaces (IRSs) constitute passive devices, which are capable of adjusting the phase shifts of their reflected signals, and hence they are suitable for passive beamforming.
    In this paper, we conceive their design with the active beamforming action of multiple-input multiple-output (MIMO) systems used at the access points (APs) for improving the beamforming gain, where both the APs and users are equipped with multiple antennas.
    Firstly, we decouple the optimization problem and design the active beamforming for a given IRS configuration.
    Then we transform the optimization problem of the IRS-based passive beamforming design into a tractable non-convex quadratically constrained quadratic program (QCQP).
    For solving the transformed problem, we give an approximate solution based on the technique of widely used semidefinite relaxation (SDR).
    We also propose a low-complexity iterative solution.
    We further prove that it can converge to a locally optimal value.
    Finally, considering the practical scenario of discrete phase shifts at the IRS, we give the quantization design for IRS elements on basis of the two solutions.
    Our simulation results demonstrate the superiority of the proposed solutions over the relevant benchmarks.
\end{abstract}
\section{Introduction}
\label{Introduction}
The fifth-generation (5G) and beyond wireless networks have a 1000-time increasing demand for network capacity.
To achieve this goal, a variety of techniques have been proposed, including millimeter-wave communications, massive multiple-input multiple-output (MIMO) and ultra-dense networks \cite{6824752}.
However, the concomitant increased energy consumption and hardware cost imposed by active devices such as antennas and radio frequency (RF) chains still make these techniques challenging for practical implementation at high frequency.
Intelligent reflecting surface (IRS) based techniques are capable of further enhancing the system's performance at a low hardware complexity and power consumption \cite{9020088,9001052,8972927}.
    IRS-assisted MIMO systems rely on passive elements for reflecting the incident signals with a particular phase shift \cite{8811733,2003.12049,1910.00959}.
    The joint design of the active baseband beamforming and passive IRS-assisted beamforming has a substantial further promise.
\subsection{Literature Review}
    Researchers have designed active and passive beamforming for a variety of IRS-assisted multiple-input single-output (MISO) systems, where a single data stream supported by a single-antenna user.
    Specifically, Wu and Zhang \cite{8811733} propose to minimize the total transmit power by applying semidefinite relaxation and alternating optimization techniques.
    They also considered realistic discrete phase shifts at the IRS \cite{8930608}, where they propose an optimal solution based on the popular branch-and-bound method as well as a suboptimal iterative solution.
    Guo \textit{et al}. \cite{8982186} propose to maximize the weighted sum-rate (WSR) of a multiuser MISO (MU-MISO) system by relying on a fractional programming technique and on the non-convex block coordinate descent (BCD) method, where both perfect and imperfect channel state information (CSI) are considered.
    Pan \textit{et al}. \cite{IRS_Hanzo} propose to maximize the WSR of IRS-assisted MISO broadcast systems.
    They adopt the classic BCD algorithm for partitioning the original optimization problem into several subproblems and then solving them alternatively.
    Nadeem \textit{et al}. \cite{9066923} propose to maximize the minimum signal-to-interference-plus-noise ratio (SINR) through optimal linear precoder (OLP) for the cases where the channel matrix between the BS and the IRS is of rank-one and of full-rank.
    Furthermore, Li \textit{et al}. \cite{1909.11314} investigate the beamforming design in wideband MISO systems for maximizing the average sum-rate over all subcarriers of MU-MISO orthogonal frequency division multiplexing (OFDM) systems by iteratively designing the actions of the IRS elements on basis of mean square error minimization.
    Zhang \textit{et al}. \cite{8990007} derive a closed-form solution for beamforming in IRS assisted single-user MISO (SU-MISO) systems, where the statistical reflected channels is considered.
    Wang \textit{et al}. \cite{wang2020intelligent} propose a joint design of active and passive precoding design algorithm for IRS assisted millimeter wave (mmWave) MISO systems, where multiple IRSs are deployed to assist the data transmission.

    Different from the above mentioned IRS-assisted MISO systems, some researchers have also studied the joint active and passive beamforming design for the MIMO systems, where multiple antennas are equipped at both the AP and the user.
    The beamforming design for IRS-assisted MIMO systems is more challenging since IRS elements and the baseband beamforming need to be properly configured for improving the system performance with multiple spatial data streams.
    Pan \textit{et al}. \cite{9090356} focus on multicell MIMO communication systems. They deal with the inter-cell interference and propose to maximize the WSR of all users by jointly optimizing the active precoding matrices at the BS and the phase shifts at the IRS.
    However, this method is too complicated for practical implementation.
	Wang \textit{et al}. \cite{wang2020joint} aim to maximize the spectrum efficiency by jointly optimizing the IRS and the hybrid precoder (combiner) at the BS (user) in IRS-assisted mmWave MIMO systems.
Joint active precoding and passive beamforming optimization is an important problem for the implemention of IRS-assisted SU-MIMO system. However, it remains non-convex and thus is hard to be solved optimally.
Regarding this, Ning \textit{et al}. \cite{9043523} first propose to reformulate it into a relaxed problem of maximizing sum-path gains, which can be effectively solved by ADMM.
Then, Zhang \textit{et al}. \cite{9110912} provide a more strict reformulation and proposed an alternating optimization (AO)-based algorithm to obtain a high-performance solution to this problem. However, the computional complexity of the AO-based solution is quite high when the number of AP or IRS antenna elements is large. 

\subsection{Contributions}
    Against this background, our main contributions are as follows:
    \begin{itemize}
    \item{In this paper, we focus on an IRS-assisted SU-MIMO system, where a multi-antenna access point (AP) serves a multi-antenna user with the aid of an IRS.
        We propose to maximize the spectrum efficiency of IRS-assisted MIMO systems, by designing the active beamforming at the AP and the passive beamforming at the IRS.
        Specifically, we decouple the optimization problem and derive the optimal baseband beamforming design for a given IRS configuration.
	   Based on the relaxed problem given in \cite{9043523}, the passive beamforming design problem is formulated as a non-convex quadratically constrained quadratic program (QCQP), subject to the constant modulus constraints imposed by IRS elements.
        }
    \item{To tackle the non-convexity of the passive beamforming design problem considered, we first give an approximate solution based on the technique of widely used semidefinite relaxation (SDR).
	   Furthermore, we derive the relationship between the objective function and each IRS element.
	Then we proprose a different solution, i.e., iterative solution, to obtain a higher-quality solution compared to ADMM-based solution. 
	   This solution, albeit yielding lower performance than AO-based solution, is appealing on account of its low conputational complexity.
        We also demonstrate the convergence of the iterative solution.
        Specifically, during the iterations, the objective function (OF) of the transformed problem is monotonically increasing, which is proved to be upper bounded by a mathematically derived value.
        Our simulation results demonstrate the superiority of the proposed iterative solution.
        }
    \item{We further consider a practical scenario, where the phase shifts of IRS elements are selected from the discrete phase set.
    We propose a pair of quantization design procedures for achieving this goal on basis of the two solutions.
    For SDR-base solution, we directly quantize the derived phase shifts according to the number of quantization bits.
    Moreover, we propose a iterative quantization method corresponding to the iteration solution.
        }
    \end{itemize}

    The remainder of this paper is organized as follows.
    In Section \ref{S2}, our system and channel models are introduced.
    In Section \ref{S3}, our problem formulation are presented.
    In Section \ref{S4}, we propose our passive beamforming for IRS-assisted MIMO systems.
    In Section \ref{S5}, numerical results are provided.
    Finally, our conclusions are drawn in Section \ref{S6}.

    \emph{Notation}:
	Lower-case and upper-case boldface letters denote vectors and matrices, respectively;
	$(\cdot)^{\rm{T}}$, $(\cdot)^{\text{H}}$, $(\cdot)^{-1}$ and $(\cdot)^{\dagger}$ represent the transpose, conjugate transpose, inverse and pseudo-inverse of a matrix, respectively;
    $\text{tr}(\cdot)$ is the trace function;
    $\| \cdot \|_{\mathrm{F}}$ denotes the Frobenius norm of a matrix;
    $\otimes$ is the Kronecker product;
    $|a|$ is the absolute value of a scalar;
    $|\mathbf{A}|$ is the determinant of a matrix;
    $\mathbf{A}_{\left[i,:\right]}$ and $\mathbf{A}_{\left[:,j\right]}$ represent the $i$-th row and $j$-th column of the matrix $\mathbf{A}$, respectively.
	Finally, $\mathbf{I}_P$ denotes the identity matrix of size $P\times P$.

	\section{System and Channel Model}\label{S2}
    In this section, we will introduce the system model and channel model of our IRS-assisted MIMO system.
	\subsection{System Model}\label{S2.1}
\begin{figure}[t]
		\center{\includegraphics[width=0.5\textwidth]{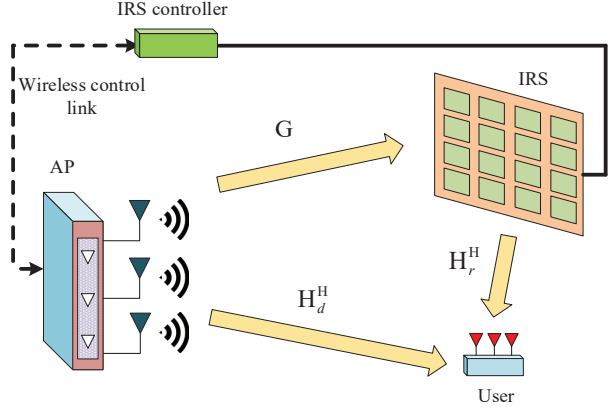}}
		\caption{Illustration of IRS-assisted MIMO systems.}
		\label{IRS}
	\end{figure}

    We consider the point-to-point MIMO system of figure \ref{IRS}, where an AP equipped with $N_t$ transmit antennas (TAs) transmits $N_s$ data streams to a user having $N_r$ receive antennas (RAs) $\left(N_s = N_r \leq N_t\right)$.
    For improving the beamforming gain during the transmission, an IRS consisting of $M$ passive elements is installed.
    We define the baseband transmit signals as $\mathbf{\mathbf{s}} = \left[ s_1,s_2,\cdots,s_{N_s} \right]^{\text{T}} \in \mathbb{C}^{N_s \times 1}$, which satisfy $\mathbb{E}\left[{\mathbf{s}\mathbf{s}^{\text{H}}}\right] = \mathbf{I}_{N_s}$.
    The AP uses baseband beamforming defined by the active beamforming matrix $\mathbf{W} \in \mathbb{C}^{N_{t} \times N_s}$.
    Then, the AP transmits the precoded signals $\mathbf{W}\mathbf{s}$ to the user through both the direct link and the reflected link.
    The channel state information (CSI) is assumed to be known at the AP \cite{8811733}.
    The direct link between the AP and the user is denoted by $\mathbf{H}_d^{\text{H}} \in \mathbb{C}^{N_r\times N_t}$.
    The reflected link is composed of three components, which are the AP-IRS channel $\mathbf{G} \in \mathbb{C}^{M\times N_t}$, IRS-induced phase shifts $\mathbf{\Theta} = \text{diag}\left( \pmb{\theta}_{\mathrm{v}} \right) \in \mathbb{C}^{M\times M}$ and the IRS-user channel $\mathbf{H}_r^{\text{H}} \in \mathbb{C}^{N_r\times M}$.
    The signals reflected by the IRS once are considered while those reflected by IRS for two and more times are ignored \cite{9043523}.
    Then, the signals $\mathbf{y} \in \mathbb{C}^{N_r\times 1}$ received at the user can be expressed as
    \begin{align}\label{y_received}
    \mathbf{y} = \sqrt{\frac{\rho}{N_s}} \left( \mathbf{H}_r^{\text{H}} \mathbf{\Theta} \mathbf{G} + \mathbf{H}_d^{\text{H}} \right) \mathbf{W} \mathbf{s} + \mathbf{n},
	\end{align}
    where $\rho$ denotes the transmit power and $\mathbf{n} \sim \mathcal{CN}\left(0, \sigma^2\mathbf{I} \right) \in \mathbb{C}^{N_s\times 1}$ represents the additive white Gaussian noise (AWGN) at the user.
    \subsection{Channel Model}\label{S2.2}
    The direct link is modeled by the Rayleigh fading channel \cite{8811733}
     \begin{align}\label{Hd}
    \mathbf{H}_d = L_0\left(d_0\right) \widetilde{\mathbf{H}}_d,
	\end{align} and the reflected links are characterized by the Rician fading model, which can be expressed by \cite{1912.10209v2,8990007}
    \begin{align}\label{Hr}
    \mathbf{H}_r = L_1\left(d_1\right) \left( \sqrt{\frac{\kappa_1}{1+\kappa_1}} \overline{\mathbf{H}}_r +  \sqrt{\frac{1}{1+\kappa_1}} \widetilde{\mathbf{H}}_r \right),
	\end{align}
    \begin{align}\label{G}
    \mathbf{G} = L_2\left(d_2\right) \left( \sqrt{\frac{\kappa_2}{1+\kappa_2}} \overline{\mathbf{G}} +  \sqrt{\frac{1}{1+\kappa_2}} \widetilde{\mathbf{G}} \right),
	\end{align}
    where $\kappa_0$, $\kappa_1$ and $\kappa_2$ are the corresponding Rician factors.
    The matrices $\overline{\mathbf{H}}_r$ and $\overline{\mathbf{G}}$ denote the line-of-sight (LoS) components of the channels.
    The matrices $\widetilde{\mathbf{H}}_d$, $\widetilde{\mathbf{H}}_r$ and $\widetilde{\mathbf{G}}$ stand for the non-LoS (NLoS) components, whose elements are independently taken from the circularly symmetric complex Gaussian (CSCG) distribution $\mathcal{CN}\left(0, 1 \right)$ \cite{8746155}.
    The path loss $L_0\left(d_0\right)$, $L_1\left(d_1\right)$ and $L_2\left(d_2\right)$ are the distance-aware attenuation, which will be further discussed in Section \ref{S5}.
    Variables $d_0$, $d_1$ and $d_2$ denote the distance between the AP and the user, the distance between the AP and the IRS, and the distance between the IRS and the user, respectively.

    We further introduce the LoS components under the assumption that the AP, the IRS and the user are equipped with uniform planar arrays (UPAs).
    The UPA departure/arrival steering vector is given by
    \begin{align}\label{a_steer}
    \mathbf{a}\left( \phi, \psi \right)& =  \left[ 1,\cdots, e^{j\frac{2\pi}{\lambda}d\left( m \mathrm{sin} \phi \mathrm{sin} \psi + n \mathrm{cos}\psi \right) } \right.\nonumber\\
    & \left.,\cdots, e^{j\frac{2\pi}{\lambda}d\left( \left( W-1 \right) \mathrm{sin} \phi \mathrm{sin} \psi + \left( H-1 \right) \mathrm{cos}\psi \right) } \right]^{\text{T}},
	\end{align}
    where $\phi$ and $\psi$ denote the azimuth and elevation angle of departure/arrival (AoD/AoA) of the signals, $\lambda$ is the wavelength, while $d$ is the distance between the adjacent antenna elements.
    The variables $m \in \{0,1,\cdots,W-1 \}$ and $n \in \{0,1,\cdots,H-1 \}$ associated with $W$ and $H$ denote the number of antennas in the horizontal and vertical directions.
    Then we can express $\overline{\mathbf{H}}_r$ and $\overline{\mathbf{G}}$ as
    \begin{align}\label{Hr_LoS}
    \overline{\mathbf{H}}_r = \mathbf{a}\left( \phi^r_1, \psi^r_1 \right)\mathbf{a}^{\text{H}}\left( \phi^t_1, \psi^t_1 \right),
	\end{align}
    \begin{align}\label{G_LoS}
    \overline{\mathbf{G}} = \mathbf{a}\left( \phi^r_2, \psi^r_2 \right)\mathbf{a}^{\text{H}}\left( \phi^t_2, \psi^t_2 \right),
	\end{align}
    where $\mathbf{a}\left( \phi^r_0, \psi^r_0 \right)$, $\mathbf{a}\left( \phi^r_1, \psi^r_1 \right)$ and $\mathbf{a}\left( \phi^r_2, \psi^r_2 \right)$ denote the arrival steering vectors, and $\mathbf{a}\left( \phi^t_0, \psi^t_0 \right)$, $\mathbf{a}\left( \phi^t_1, \psi^t_1 \right)$ and $\mathbf{a}\left( \phi^t_2, \psi^t_2 \right)$ represent the departure steering vectors.

    \section{Problem Formulation}\label{S3}
    In this section, we focus our attention on maximizing the spectrum efficiency of IRS-assisted MIMO systems.
    Then we transform our intractable problem into a non-convex QCQP for solving it.

    The optimization problem for maximizing the spectrum efficiency can be formulated as
    \begin{subequations}\label{ObjSU}
    \begin{align}
    \left( \text{P}1 \right): \ \max_{\mathbf{\Theta},\mathbf{\mathbf{W}} } \quad & \mathrm{log}_2 \left(\left| \mathbf{I}_{N_r} + \frac{\rho}{N_s\sigma^2}  \left( \mathbf{H}_r^{\text{H}} \mathbf{\Theta} \mathbf{G} + \mathbf{H}_d^{\text{H}}  \right)\cdot \right.\right. \nonumber\\
    &\left.\left. \mathbf{W}  \mathbf{W}^{\text{H}} \left( \mathbf{H}_r^{\text{H}} \mathbf{\Theta} \mathbf{G} + \mathbf{H}_d^{\text{H}}  \right)^{\text{H}} \right|\right) \\
    s.t. \ \quad   & \quad \left| \mathbf{\Theta}_{m,m} \right| = 1, \forall m = 1,2,\cdots,M. \\
    & \quad \left\| \mathbf{W} \right\|_{\rm F}^2 = N_s.
    \end{align}
    \end{subequations}
    The problem (P1) is a non-convex one subject to the constant-modulus constraint of IRS elements.
    Following the conclusions of \cite{Shen}, we approximately transform the problem (P1) under the high-SNR assumption as
    \begin{subequations}\label{ObjSU_v2}
    \begin{align}
    \left( \text{P}2 \right): \quad \max_{\mathbf{\Theta}, \mathbf{W} } \quad & \left\Vert \left( \mathbf{H}_r^{\text{H}} \mathbf{\Theta} \mathbf{G} + \mathbf{H}_d^{\text{H}} \right) \mathbf{W} \right\Vert_{\text{F}}^2 \label{ObjSU_v2a} \\
    s.t. \quad & \left| \mathbf{\Theta}_{m,m} \right| = 1, \forall m = 1,2,\cdots,M. \\
    \quad & \left\| \mathbf{W} \right\|_{\rm F}^2 = N_s.
    \end{align}
    \end{subequations}

    According to the problem (P2), we have to design both the active beamforming $\mathbf{W}$ and the phase shifts $\mathbf{\Theta}$ for maximizing the spectrum efficiency.
    Observe in \eqref{ObjSU_v2a} that $\mathbf{W}$ and $\mathbf{\Theta}$ are coupled in the OF.
    For decoupling the two variables, we adopt the optimal baseband beamforming matrix for a given IRS configuration $\mathbf{\Theta}$ as $\mathbf{\mathbf{W}}_{\mathrm{opt}} = \mathbf{V}_{\mathrm{eff}\left[ :, 1:N_s \right]}$ \cite{JSAC_XGao_EnergyEfficient,8284058}
    where $\mathbf{H}_{\mathrm{eff}} = \mathbf{H}_r^{\text{H}} \mathbf{\Theta} \mathbf{G} + \mathbf{H}_d^{\text{H}} $ is the effective channel and $\mathbf{V}_{\mathrm{eff}}$ consists of the right singular vectors of $\mathbf{H}_{\mathrm{eff}}$.

    Then, upon adopting the near-optimal baseband beamforming matrix, we further transform \eqref{ObjSU_v2a} under the assumption $N_s = N_r \leq N_t$ as
    \begin{align}\label{CS_ieq}
    & \left\| \mathbf{H}_{\mathrm{eff}} \mathbf{W}_{\rm opt} \right\|_{\rm F}^2 \nonumber \\
    = & \rm{Tr}\left( \mathbf{H}_{\mathrm{eff}} \mathbf{W}_{\rm opt} \mathbf{W}_{\rm opt}^{\text{H}} \mathbf{H}_{\mathrm{eff}}^{\text{H}}\right) \nonumber \\
    = & \rm{Tr}\left( \mathbf{U}_{\mathrm{eff}} \mathbf{\Sigma}_{\mathrm{eff}} \mathbf{V}_{\mathrm{eff}}^{\text{H}} \mathbf{W}_{\rm opt} \mathbf{W}_{\rm opt}^{\text{H}} \left( \mathbf{U}_{\mathrm{eff}} \mathbf{\Sigma}_{\mathrm{eff}} \mathbf{V}_{\mathrm{eff}}^{\text{H}} \right)^{\text{H}} \right)\nonumber \\
    \overset{\left(a\right)}{=} & \rm{Tr}\left( \mathbf{U}_{\mathrm{eff}} \begin{bmatrix} \overline{\mathbf{\Sigma}}_{\mathrm{eff}} &  \mathbf{0}_{N_r \times N_q} \end{bmatrix}   \right. \nonumber \\
    & \left. \begin{bmatrix} \mathbf{I}_{N_r} &  \mathbf{0}_{N_r \times N_q} \\ \mathbf{0}_{N_q \times N_r} & \mathbf{0}_{N_q \times N_q} \end{bmatrix} \begin{bmatrix}\overline{\mathbf{\Sigma}}_{\mathrm{eff}}^{\rm H} \\  \mathbf{0}_{N_q \times N_r} \end{bmatrix}  \mathbf{U}_{\mathrm{eff}}^{\text{H}} \right)\nonumber \\
    \overset{\left(b\right)}{=} & \rm{Tr}\left( \overline{\mathbf{\Sigma}}_{\mathrm{eff}}^2 \right)\nonumber \\
    = & \rm{Tr}\left( \mathbf{\Sigma}_{\mathrm{eff}} \mathbf{\Sigma}_{\mathrm{eff}}^{\rm H}  \right)\nonumber \\
    = & \left\| \mathbf{H}_{\mathrm{eff}} \right\|_{\rm F}^2,
    \end{align}
    where $\overline{\mathbf{\Sigma}}_{\mathrm{eff}}$ is a diagonal matrix with the diagonal elements being the singular values of $\mathbf{H}_{\mathrm{eff}}$, parameter $N_q$ is defined by $N_q = N_t-N_r$, $\left(a\right)$ is due to the fact that  $\text{rank}\left(\mathbf{H}_{\mathrm{eff}}\right) = N_r$ and $ \mathbf{V}_{\mathrm{eff}}^{\text{H}} \mathbf{W}_{\rm opt} = \begin{bmatrix} \mathbf{I}_{N_r} \\ \mathbf{0}_{N_q \times N_r} \end{bmatrix} $, $\left(b\right)$ holds for $\rm{Tr}\left( \mathbf{X}\mathbf{Y} \right) = \rm{Tr}\left( \mathbf{Y}\mathbf{X} \right)$.
    In this way, we decouple the design of the active and passive beamforming matrices and transform the problem (P2) as
    \begin{subequations}\label{ObjSU_v3}
    \begin{align}
    \left( \text{P}3 \right): \quad \max_{\mathbf{\Theta} } \quad & \left\Vert \mathbf{H}_r^{\text{H}} \mathbf{\Theta} \mathbf{G} + \mathbf{H}_d^{\text{H}} \right\Vert_{\text{F}}^2 \label{ObjSU_v3a} \\
    s.t. \quad & \left| \mathbf{\Theta}_{m,m} \right| = 1, \forall m = 1,2,\cdots,M.
    \end{align}
    \end{subequations}
    Afterwards, we characterize the OF \eqref{ObjSU_v3a} in another form for decoupling the IRS elements from the channel matrices.
    Specifically, we rewrite \eqref{ObjSU_v3a} as
    \begin{align}\label{trans}
    & \|\mathbf{H}_r^{\text{H}} \mathbf{\Theta} \mathbf{G} + \mathbf{H}_d^{\text{H}} \|_{\text{F}}^2 \nonumber \\
    & = \mathrm{vec}^{\text{H}}\left( \mathbf{H}_r^{\text{H}} \mathbf{\Theta} \mathbf{G} + \mathbf{H}_d^{\text{H}} \right) \mathrm{vec}\left( \mathbf{H}_r^{\text{H}} \mathbf{\Theta} \mathbf{G} + \mathbf{H}_d^{\text{H}} \right).
	\end{align}
    By adopting the essential equation $\mathrm{vec}\left( \mathbf{A}\mathbf{B}\mathbf{C} \right) = \left( \mathbf{C}^{\mathrm{T}} \otimes \mathbf{A} \right) \mathrm{vec}\left( \mathbf{B} \right)$,
    we can further rewrite (\ref{trans}) as
    \begin{align}\label{trans_v2}
    & \mathrm{vec}^{\text{H}}\left( \mathbf{H}_r^{\text{H}} \mathbf{\Theta} \mathbf{G} + \mathbf{H}_d^{\text{H}} \right) \mathrm{vec}\left( \mathbf{H}_r^{\text{H}} \mathbf{\Theta} \mathbf{G} + \mathbf{H}_d^{\text{H}} \right) \nonumber \\
    & = \mathrm{vec}^{\text{H}}\left( \mathbf{\Theta} \right) \left( \mathbf{G}^{\text{T}} \otimes \mathbf{H}_r^{\text{H}} \right)^{\text{H}} \left( \mathbf{G}^{\text{T}} \otimes \mathbf{H}_r^{\text{H}} \right) \mathrm{vec}\left( \mathbf{\Theta} \right) \nonumber \\
    & \quad \ + \mathrm{vec}^{\text{H}}\left( \mathbf{\Theta} \right) \left( \mathbf{G}^{\text{T}} \otimes \mathbf{H}_r^{\text{H}} \right)^{\text{H}} \mathrm{vec}\left( \mathbf{H}_d^{\text{H}} \right) \nonumber \\
    & \quad \ + \mathrm{vec}^{\text{H}}\left( \mathbf{H}_d^{\text{H}} \right) \left( \mathbf{G}^{\text{T}} \otimes \mathbf{H}_r^{\text{H}} \right) \mathrm{vec}\left( \mathbf{\Theta} \right) \nonumber \\
    & \quad \ + \mathrm{vec}^{\text{H}}\left( \mathbf{H}_d^{\text{H}} \right) \mathrm{vec}\left( \mathbf{H}_d^{\text{H}} \right).
	\end{align}
    Then we simplify the expression (\ref{trans_v2}).
    The vector $\mathrm{vec}\left( \mathbf{\Theta} \right)$ can be equivalently expressed by $\mathrm{vec}\left( \mathbf{\Theta} \right) =
    \begin{bmatrix}
    \pmb{\theta}_{1}^{\text{H}}&
    \pmb{\theta}_{2}^{\text{H}}&
    \cdots &
    \pmb{\theta}_{M}^{\text{H}}
    \end{bmatrix}^{\text{H}}$,
    where $\pmb{\theta}_{m} = e^{j\theta_m}\mathbf{I}_{M\left[ :,m \right]}$.
    We observe that there are $M\left( M - 1 \right)$ zero elements in $\mathrm{vec}\left( \mathbf{\Theta} \right)$ and we define the location set of the nonzero elements as $\mathcal{M}_{\mathrm{nz}}$.
    Then we can further reduce the sizes of vectors and matrices in (\ref{trans_v2}).
    We define the coupled reflected channel matrix as $\mathbf{G}_{c} = \left( \mathbf{G}^{\text{T}} \otimes \mathbf{H}_r^{\text{H}}\right)_{\left[ :, j \in \mathcal{M}_{\mathrm{nz}} \right]}$
    and the vectorized direct channel matrix as $\mathbf{h}_{\mathrm{v}} = \mathrm{vec}\left( \mathbf{H}_d^{\text{H}} \right)$.
    It may be observed that the constant term $\mathbf{h}_{\mathrm{v}}^{\text{H}}\mathbf{h}_{\mathrm{v}}$ is not related to the IRS elements, which can be ignored.
    Thus, recalling $\pmb{\theta}_{\mathrm{v}}$ defined in Section \ref{S2.1}, we can equivalently transform the problem (P3) as
    \begin{subequations}\label{ObjSU_v4}
    \begin{align}
    \left( \text{P}4 \right): \quad \max_{ \pmb{\theta} } \quad & \pmb{\theta}_{\mathrm{v}}^{\text{H}} \mathbf{G}_{c}^{\text{H}} \mathbf{G}_{c} \pmb{\theta}_{\mathrm{v}} + \pmb{\theta}_{\mathrm{v}}^{\text{H}} \mathbf{G}_{c}^{\text{H}} \mathbf{h}_{\mathrm{v}} + \mathbf{h}_{\mathrm{v}}^{\text{H}}\mathbf{G}_{c} \pmb{\theta}_{\mathrm{v}} \\
    s.t. \quad & \left| \pmb{\theta}_{\mathrm{v},m} \right| = 1, \forall m = 1,2,\cdots,M.
    \end{align}
    \end{subequations}
    Note that although the problem (P4) is in the form of a well-known non-convex QCQP, the problem transformation from the (P1) to (P4) is specific for IRS-assisted MIMO systems.
    In the next section, we will give an approximate SDR-based solution and propose a low-complexity iterative solution.
    \section{Passive Beamforming for IRS-assisted MIMO systems}\label{S4}
    In this section, we will first give an approximate solution for the problem (P4) based on SDR.
    Then, we further propose a low-complexity iterative solution for approaching the optimal spectrum efficiency.
    \subsection{SDR-based Solution}\label{S4.1}
    By introducing an auxiliary variable $t$, the problem (P4) can be reformulated as a homogeneous QCQP \cite{8811733,1912.10209v2}, which can be expressed as
    \begin{subequations}\label{ObjSU_SDRv1}
    \begin{align}
    \left( \text{P}5 \right): \quad  \max_{\tilde{\pmb{\theta}}_{\mathrm{v}} } \quad & \tilde{\pmb{\theta}}_{\mathrm{v}}^{\text{H}} \mathbf{R} \tilde{\pmb{\theta}}_{\mathrm{v}} \\
    s.t. \quad &\left| \tilde{\pmb{\theta}}_{\mathrm{v},m} \right| = 1, \forall m = 1,2,\cdots,M+1,
    \end{align}
    \end{subequations}
    where $\tilde{\pmb{\theta}}_{\mathrm{v}} = \begin{bmatrix}
    \pmb{\theta}_{\mathrm{v}} \\
    t
    \end{bmatrix}$,
    $\mathbf{R} = \begin{bmatrix}
    \mathbf{G}_{c}^{\text{H}}\mathbf{G}_{c} & \mathbf{G}_{c}^{\text{H}}\mathbf{h}_{\mathrm{v}}\\
    \mathbf{h}_{\mathrm{v}}^{\text{H}}\mathbf{G}_{c} & 0
    \end{bmatrix}$.
    We define $\mathbf{V} = \tilde{\pmb{\theta}}_{\mathrm{v}}\tilde{\pmb{\theta}}_{\mathrm{v}}^{\text{H}}$, which satisfies $\mathbf{V}\succeq \mathbf{0}$ and $\text{rank}\left( \mathbf{V} \right) = 1$.
    Then we utilize the SDR for relaxing the rank-one constraint and transform the problem (P5) as
    \begin{subequations}\label{ObjSU_SDRv2}
    \begin{align}
    \left( \text{P}6 \right): \quad  \max_{\tilde{\pmb{\theta}}_{\mathrm{v}} } \quad  & \text{tr}\left( \mathbf{R}\mathbf{V} \right) \\
    s.t. \quad  & \mathbf{V}_{m,m} = 1, \forall m = 1,2,\cdots,M+1, \\
    & \mathbf{V} \succeq 0.
    \end{align}
    \end{subequations}
    We observe that the problem (P6) is a standard convex semidefinite program (SDP), hence we can solve it by existing convex optimization solvers such as CVX for acquiring the optimal solution \cite{timmurphy.org}.
    Due to the fact that the solution may not satisfy the rank-one constraint, the following Gaussian randomization procedure is adopted for producing a high-quality rank-one solution \cite{5447068}.
    Specifically, we first derive the eigenvalue decomposition of $\mathbf{V}$ as $\mathbf{V} = \mathbf{U}_s \mathbf{\Sigma}_s \mathbf{U}_s^{\text{H}}$.
    Then we obtain the modified solution to (P5) as $\tilde{\pmb{\theta}}_{\mathrm{v}} = \mathbf{U}_s \mathbf{\Sigma}_s^{\frac{1}{2}}\mathbf{r}$.
    The vector $\mathbf{r}\sim \mathcal{CN}\left(0, \mathbf{I}_{M+1} \right) \in \mathbb{C}^{\left(M+1\right)\times 1}$ is selected from a large number of random generalized CSCG vectors for maximizing the OF value of (P5).
    Finally, the solution to problem (P4) is acquired by $\pmb{\theta}_{\mathrm{v}} = e^{j\arg \left( \tilde{\pmb{\theta}}_{\mathrm{v}\left[ 1:M\right]}/\tilde{\pmb{\theta}}_{\mathrm{v}\left[M+1\right]} \right)}$,
    which satisfies the constant-modulus constraint.

    We summarize the SDR-based solution in \textbf{Algorithm \ref{alg:algorithm1}}.
    Specifically, the CVX solution acquired in step 1 is followed by Gaussian randomization procedure in step 2.
    The final solution is obtained in step 3.

\begin{algorithm}[h]
\caption{SDR-based solution}
\label{alg:algorithm1}
    \begin{algorithmic}[1]
        \REQUIRE ~~\\
        $\mathbf{R}$;\\
        \ENSURE
        \STATE Solve Problem (P6) through CVX and acquire $\mathbf{V}$;
        \STATE Adopt Gaussian randomization procedure for getting $\tilde{\pmb{\theta}}_{\mathrm{v}} = \mathbf{U}_s \mathbf{\Sigma}_s^{\frac{1}{2}}\mathbf{r}$.
        \STATE Acquire the IRS elements as $\pmb{\theta}_{\mathrm{v}} = e^{j\arg \left( \tilde{\pmb{\theta}}_{\mathrm{v},\left[ 1:M\right]}/\tilde{\pmb{\theta}}_{\mathrm{v},M+1} \right)}$.
    \end{algorithmic}
\end{algorithm}
    \subsection{Iterative Solution}\label{S4.2}
    For mitigating the complexity of the SDR-based solution, we now propose a low-complexity iterative solution.
    Specifically, we first equivalently transform the problem (P4) as
    \begin{subequations}\label{ObjSU_CIMv1}
    \begin{align}
    \left( \text{P}7 \right): \quad  \max_{\overline{\pmb{\theta}}_{\mathrm{v}} } \quad & \overline{\pmb{\theta}}_{\mathrm{v}}^{\text{H}} \mathbf{R} \overline{\pmb{\theta}}_{\mathrm{v}} \label{ObjSU_CIMv1a} \\
    s.t. \quad &\left| \overline{\pmb{\theta}}_{\mathrm{v},m} \right| = 1, \forall m = 1,2,\cdots,M+1,
    \end{align}
    \end{subequations}
    where $\overline{\pmb{\theta}}_{\mathrm{v}} = \begin{bmatrix}
    \pmb{\theta}_{\mathrm{v}} \\
    1
    \end{bmatrix}$.
    We propose to design the phase shifts of the IRS elements in a one-by-one manner during the iterations.
    In order to update the phase shifts, we decompose the OF \eqref{ObjSU_CIMv1a} into an element-wise form as
    \begin{align}\label{decomp_CIM}
    & \quad \ \overline{\pmb{\theta}}_{\mathrm{v}}^{\text{H}} \mathbf{R} \overline{\pmb{\theta}}_{\mathrm{v}}\nonumber \\
    & = e^{-j\theta_{m}}\sum_{j\neq m}^{M+1} e^{j\theta_{j}} \mathbf{R}_{\left[m,j\right]} + e^{j\theta_{m}}\sum_{i\neq m}^{M+1} e^{-j\theta_{i}} \mathbf{R}_{\left[i,m\right]}\nonumber  \\
    & + \sum_{i\neq m}^{M+1} e^{-j\theta_{i}} \sum_{j\neq m}^{M+1} e^{j\theta_{j}} \mathbf{R}_{\left[i,j\right]} + \mathbf{R}_{\left[m,m\right]}\nonumber \\
    & \overset{\left(a\right)}{=} 2 \mathfrak{Re}\left(e^{j\theta_{m}}\sum_{i\neq m}^{M+1} e^{-j\theta_{i}} \mathbf{R}_{\left[i,m\right]} \right) \nonumber\\
    & + \sum_{i\neq m}^{M+1} e^{-j\theta_{i}} \sum_{j\neq m}^{M+1} e^{j\theta_{j}} \mathbf{R}_{\left[i,j\right]}  + \mathbf{R}_{\left[m,m\right]},
    \end{align}
    where (a) holds due to the fact that $\mathbf{R}$ is a Hermitian matrix.
    Since only the first term in (\ref{decomp_CIM}) contains $e^{j\theta_m}$, the optimal phase shift of the $m$-th IRS element can be obtained as
    \begin{align}\label{theta_opt}
    \theta_m^{\star} = \mathrm{angle}\left(\mathrm{conj} \left( \sum_{i\neq m}^{M+1} e^{-j\theta_{i}} \mathbf{R}_{\left[i,m\right]} \right) \right).
    \end{align}
    We successively design the phase shifts for all IRS elements in the order from $m=1$ to $m=M$.
    After we obtain the phase shifts $\left\{ \theta_m^{\star} \right\}_{m=1}^M$ for all elements, we will repeat the process until convergence is attained.
    Note that the OF value in problem (P7) is non-decreasing during the iterations.
    Furthermore, the upper bound of (\ref{ObjSU_CIMv1}a) can be derived as $\overline{\pmb{\theta}}_{\mathrm{v}}^{\text{H}} \mathbf{R} \overline{\pmb{\theta}}_{\mathrm{v}} \leq \left( M+1 \right)\lambda_{\max}\left( \mathbf{R} \right)$ \cite{8930608},
    where $\lambda_{\max}\left( \mathbf{R} \right)$ denotes the maximum eigenvalue of $\mathbf{R}$.
    Therefore, the proposed iterative solution may converge to a locally optimal solution.

    Now we will give the complexity analysis of the iterative solution.
    According to (\ref{theta_opt}), the calculation of phases has the computational complexity order of $\mathcal{O}\left( M \right)$.
    The process has to be repeated $K$ times for convergency.
    Hence the order of complexity for passive beamforming design is $\mathcal{O}\left( M^2K \right)$.

    We summarize the proposed iterative solution in \textbf{Algorithm \ref{alg:algorithm2}}.
    In step 2-4, all IRS elements are updated, which is repeated $K$ times for convergence.
\begin{algorithm}[h]
\caption{Iterative solution}
\label{alg:algorithm2}
    \begin{algorithmic}[1]
        \REQUIRE ~~\\
        $\mathbf{R}$, iteration numer $K$;\\
        \ENSURE
        \FOR{$k = 1 : K$}
        \FOR{$m = 1 : M$}
        \STATE Calculate and update the optimal phase shift $\theta_m^{\star}$ according to (\ref{theta_opt});
        \ENDFOR
        \ENDFOR
    \end{algorithmic}
\end{algorithm}

    \subsection{Extend to Discrete Phase Shifts scenario}\label{S4.4}
    In this section, we consider the impact of discrete phase shifts at the IRS.
    We consider the phase shifts of IRS elements as $\mathbf{\Theta}_{m,m} \in \mathcal{F}$, where $\mathcal{F} = \left\{ -\pi + \frac{2\pi}{2^B}, -\pi + \frac{2\pi 2}{2^B}, \cdots, \pi \right\}$ denotes the discrete phase set.
    The variable $B$ is the number of quantization bits of phase shifts.

    We propose a pair of quantization design procedures.
    For SDR-based solution, we directly quantize the derived phase shifts as $\pmb{\theta}_{\mathrm{v}} = e^{j \mathcal{Q}\left( \arg \left( \tilde{\pmb{\theta}}_{\mathrm{v}\left[ 1:M\right]}/\tilde{\pmb{\theta}}_{\mathrm{v}\left[M+1\right]} \right) \right)}$,    where $\mathcal{Q}\left( \cdot \right)$ denotes the quantization function, which quantizes the complex unit-norm variables in a vector to the nearest point in the set $\mathcal{F}$.

    We further propose an iterative quantization procedure, which is acquired by slightly modifying the iterative solution for the discrete phase shifts scenario.
    Specifically, during the design for the $m$-th IRS element, our goal is to maximize the OF (\ref{ObjSU_CIMv1}a).
    This is equivalent to minimizing the angle between the discrete phase shift $\theta_m^{\rm D}$ and $\theta_m^{\star}$ derived by (\ref{theta_opt}) on the complex plane, where $\theta_m^{\rm D}$ is selected from $\mathcal{F}$.
    Therefore, the discrete phase shift during the design for the $m$-th element is acquired by $\theta_m^{\rm D} = \mathcal{Q}\left( \theta_m^{\star} \right)$.

    \section{Numerical Results}\label{S5}

    In this section, we provide our simulation results characterizing the proposed systems.
    The AP has $N_t = 4\times4$ TAs and the user has $N_r=4\times4$ RAs.
    The number of data streams is set to $N_s = 16$.
    By contrast, the Rician factors of the reflected channels are set to $\kappa_1 \rightarrow \infty$ and $\kappa_2 = 10$dB \cite{2002.03744v2,1912.10209v2}.
    The azimuth AoD $\phi^{\mathrm{r}}_0$, $\phi^{\mathrm{r}}_1$, $\phi^{\mathrm{r}}_2$ and AoA $\phi^{\mathrm{t}}_0$, $\phi^{\mathrm{t}}_1$, $\phi^{\mathrm{t}}_2$ are uniformly distributed in the interval $\left( -\pi,\pi \right]$.
    The elevation AoD $\psi^{\mathrm{r}}_0$, $\psi^{\mathrm{r}}_0$, $\psi^{\mathrm{r}}_0$ and AoA $\psi^{\mathrm{t}}_0$, $\psi^{\mathrm{t}}_1$, $\psi^{\mathrm{t}}_2$ are uniformly distributed in the interval $\left( -\frac{\pi}{2},\frac{\pi}{2} \right]$ \cite{TWC_AAhmed_LimitedHybridPrecoding}.

 	\begin{figure}[t]
		\center{\includegraphics[width=0.5\textwidth]{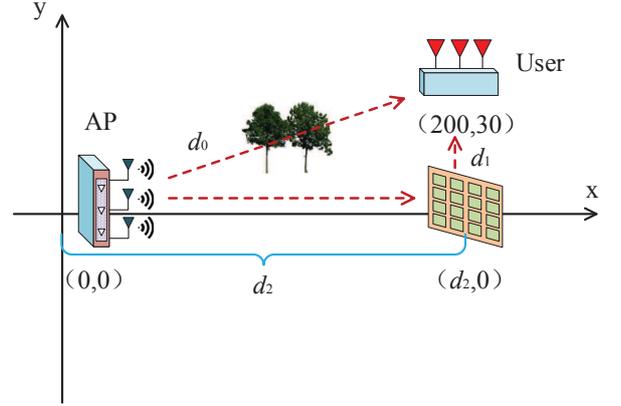}}
		\caption{Illustration of the simulated MIMO communication scenario.}
		\label{Simulation}
	\end{figure}
    The practical scenario considered for simulation is illustrated in figure \ref{Simulation}.
    The AP is located in $\left( 0,0 \right)$.
    The multi-antenna user is located in $\left( 200,30 \right)$.
    The distance between the AP and the IRS is $d_2$ and hence the location of the IRS is $\left( d_2,0 \right)$.
    The unit of the distance is meter.
    The path loss of the direct link is calculated by 32.6 + 36.7$\log d_0$ dB and that for the reflected link is 35.6 + 22.0$\log \left(d_1 + d_2\right)$ dB \cite{8982186}.
    The noise power spectral density is set to -170 dBm/Hz and transmission bandwidth is set to 180 kHz \cite{8982186}.

\begin{figure}[t]
		\center{\includegraphics[width=0.5\textwidth]{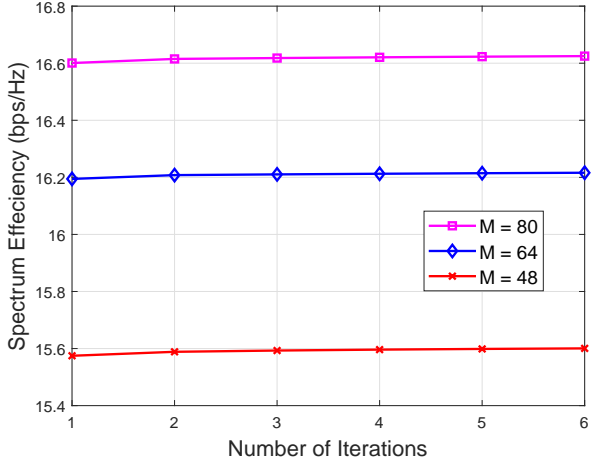}}
		\caption{Convergence of the proposed iterative algorithm.}
		\label{Iteration}
	\end{figure}

    We first demonstrate the convergence of the proposed iterative solution.
    We consider the cases, where the number of IRS elements is set to $M = 48$, $M = 64$ and $M = 80$.
    The transmit power is set to $\rho = 30$dBm and the distance is set to $200$m.
    As shown in figure \ref{Iteration}, at most 3 iterations are required for convergence, which indicates that our iterative solution has a low complexity.
    Therefore, in the following simulations, the number of iterations is set to $5$.
 	\begin{figure}[t]
		\center{\includegraphics[width=0.5\textwidth]{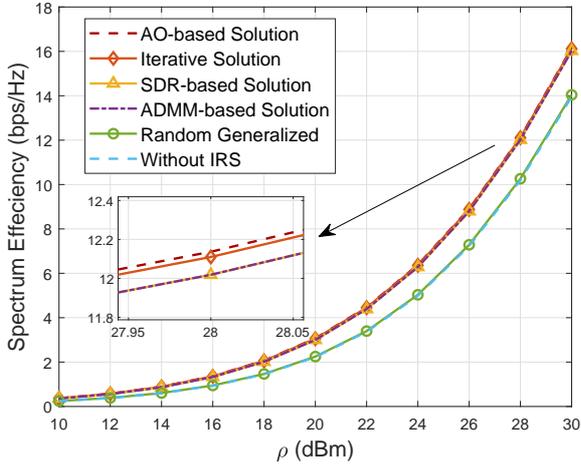}}
		\caption{Spectrum efficiency versus transmit power $\rho$.}
		\label{IRS_method}
	\end{figure}
 	\begin{figure}[t]
		\center{\includegraphics[width=0.5\textwidth]{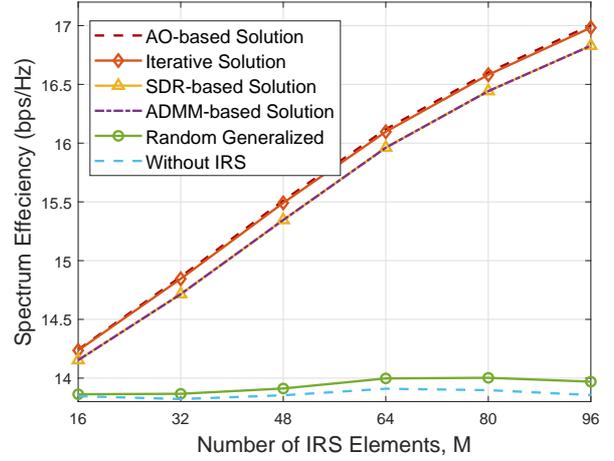}}
		\caption{Spectrum efficiency versus the number of IRS elements $M$.}
		\label{N_IRS}
	\end{figure}
 	\begin{figure}[t]
		\center{\includegraphics[width=0.5\textwidth]{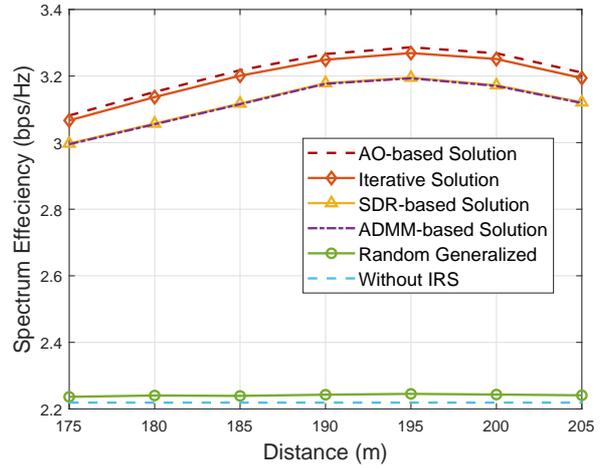}}
		\caption{Spectrum efficiency versus distance $d_2$.}
		\label{SE_D}
	\end{figure}
 	\begin{figure}[t]
		\center{\includegraphics[width=0.5\textwidth]{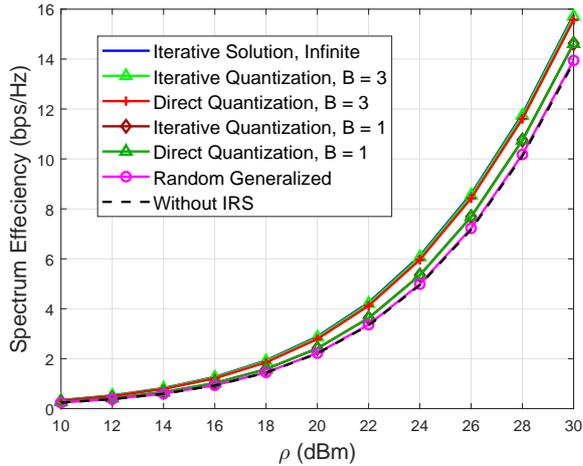}}
		\caption{Spectrum efficiency versus transmit power $\rho$ with different numbers of quantization bits $B$.}
		\label{SE_bit}
	\end{figure}

    In figure \ref{IRS_method}, we show the spectrum efficiency of our solutions and benchmarks.
    The number of IRS elements is set to $M = 64$, the distance $d_2$ is set to 200 and the transmit power ranges from $10$dBm to $30$dBm.
    We consider the following three benchmarks for comparison:
    1) ADMM-based solution \cite{9043523}.
    2) Random Generalized: $M$ phase shifts $\pmb{\theta}_{\mathrm{v}}$ are randomly selected from a uniform distribution $\left( -\pi, \pi \right]$.
    3) Without IRS: the conventional fully digital MIMO benchmark.
    4) AO-based solution: the AO-based solution proposed in \cite{9110912}.
    We observe that our proposed iterative solution imposes much lower complexity whilst having higher spectrum efficiency compared with SDR-based solution.
    This is because the SDR-based solution only extract the optimal phases and the modulus of IRS element are set to 1 compulsively.
    By contrast, in the proposed iterative solution, the phase shift of each IRS element is obtained in a one-by-one manner by maximizing the OF (18a).
During the iteration, the phase shift of each IRS element is derived under the constant-modulus constraint.
The iterative solution does not need the process of phase extraction, which is used in the SDR-based solution and inevitably deteriorates the spectrum efficiency.
Therefore, the proposed iterative solution behaves better than the SDR-based solution.
    Our proposed iterative solution shows a similar spectrum efficiency performance compared to AO-based solution, whilst having a lower computational complexity.
    We can also observe that the iterative solution has a slightly higher spectrum efficiency than ADMM-based solution, and the computational complexity is lower when the number of IRS elements is large.
    Compared to the traditional MIMO systems, the spectrum efficiency gain of random generalized solution is marginal.
    However, the IRS-assisted MIMO systems have an increased spectrum efficiency after adopting the proposed low-complexity iterative solution.

    In figure \ref{N_IRS}, we investigate the spectrum efficiency of IRS-assisted MIMO systems against the number of IRS elements.
    In the simulations, we set the transmit power to $\rho = 30$dBm and the distance $d_2 = 200$.
    Observe that the proposed iterative solution shows a slightly better spectrum efficiency than the SDR-based solution.
    This is because that the SDR-based solution ameliorated by the Gaussian randomization procedure is a suboptimal solution \cite{8811733}.
    The proposed low-complexity iterative solution shows a similar spectrum efficiency performance compared to AO-based solution and has a slightly higher spectrum efficiency than ADMM-based solution.
    Furthermore, we observe that the spectrum efficiency of IRS-assisted MIMO systems increases with $M$, because increasing $M$ leads to an improved beamforming gain.

    figure \ref{SE_D} illustrates the relationship between the spectrum efficiency and the distance between the AP and the IRS.
    The number of IRS elements is set to $M = 64$.
    The transmit power is set to $\rho$=20dBm, while moving the IRS from (175, 0) to (205, 0).
    It is observed that when we increase $d_2$ from 175 to 205, the spectrum efficiency first increases, and then decreases.
    The spectrum efficiency reaches the peak when $d_2$ is about 195.
    This is because the path loss of the reflected link is the product of the path losses of the channel $\mathbf{G}$ and the channel $\mathbf{H}_r$, which decreases when $d_2$ increases from 175 to 195 and then increases as $d_2$ continues to increase from 195 to 205.

    In figure \ref{SE_bit}, we study the relationship between the spectrum efficiency and the resolution of phase shifts $B$.
    The number of IRS elements is set to $M = 64$, the distance $d_2$ is set to 200.
    In addition to the ideal case where the phase shifts have infinite resolution, we set two finite resolution cases for comparison, including $B = 3$bits and $B = 1$bit.
    We observe that the 3-bit resolution phase shifts can provide a near-optimal spectrum efficiency performance compared with the infinite-resolution one.
    Even when the resolution of phase shifts is set to 1 bit, the spectrum efficiency is higher than that of two counterparts, which are random generalized phase shifts systems and the traditional MIMO systems.
    Besides, we also observe that the iterative quantization procedure can achieve a higher spectrum efficiency compared to the direct quantization.
    \section{Conclusion}\label{S6}
    In this paper, we studied an IRS-assisted point-to-point MIMO system, where both the AP and the user are equipped with multiple antennas.
    We maximized the spectrum efficiency by designing the active baseband beamforming at the AP and the passive beamforming at the IRS.
    Specifically, we decoupled the design problem and adopted the optimal baseband beamforming in conjunction with a given IRS configuration.
    Then, we transformed the optimization problem of the IRS elements into a non-convex QCQP by decoupling the IRS elements from the channel matrices.
    A low-complexity iterative solution was proposed for passive beamforming.
    Our simulation results demonstrated the superiority of our proposed iterative solution over the benchmarks.
\section*{ACKNOWLEDGEMENT}
\label{ACKNOWLEDGEMENT}
This work was supported in part by the the National Key Research and Development Program of China under No. 2019YFB1803200, and by the National Natural Science Foundation of China (NSFC) under Grant 61620106001 and 61901034.
\bibliographystyle{gbt7714-numerical}
\bibliography{myref}

\biographies
\begin{CCJNLbiography}{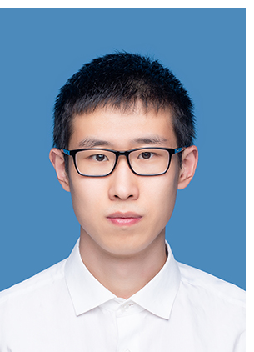}{Chenghao Feng}
received the B.E. degree from the Beijing Institute of Technology, Beijing, China, in 2017, where he is currently pursuing the Ph.D. degree with the School of Information and Electronics. His current research interests include massive MIMO, mmWave/THz communications, energy-efficient communications, intelligent reflecting surface and networks.
\end{CCJNLbiography}

\begin{CCJNLbiography}{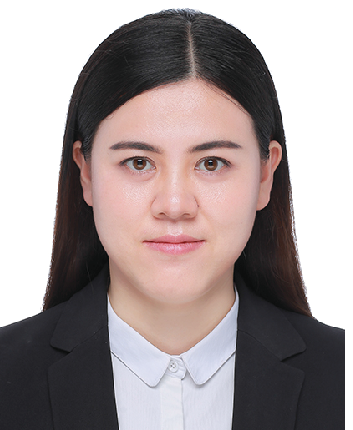}{Wenqian Shen}
received the B.S. degree from Xi'an Jiaotong University, Shaanxi, China in 2013 and the Ph.D. degree from Tsinghua University, Beijing, China. She is currently an associate professor with the School of Information and Electronics, Beijing Institute of Technology, Beijing, China in 2018. Her research interests include massive MIMO and mmWave/THz communications. She has published several journal and conference papers in IEEE Transaction on Signal Processing, IEEE Transaction on Communications, IEEE Transaction on Vehicular Technology, IEEE ICC, etc. She has won the IEEE Best Paper Award at the IEEE ICC 2017.
\end{CCJNLbiography}

\begin{CCJNLbiography}{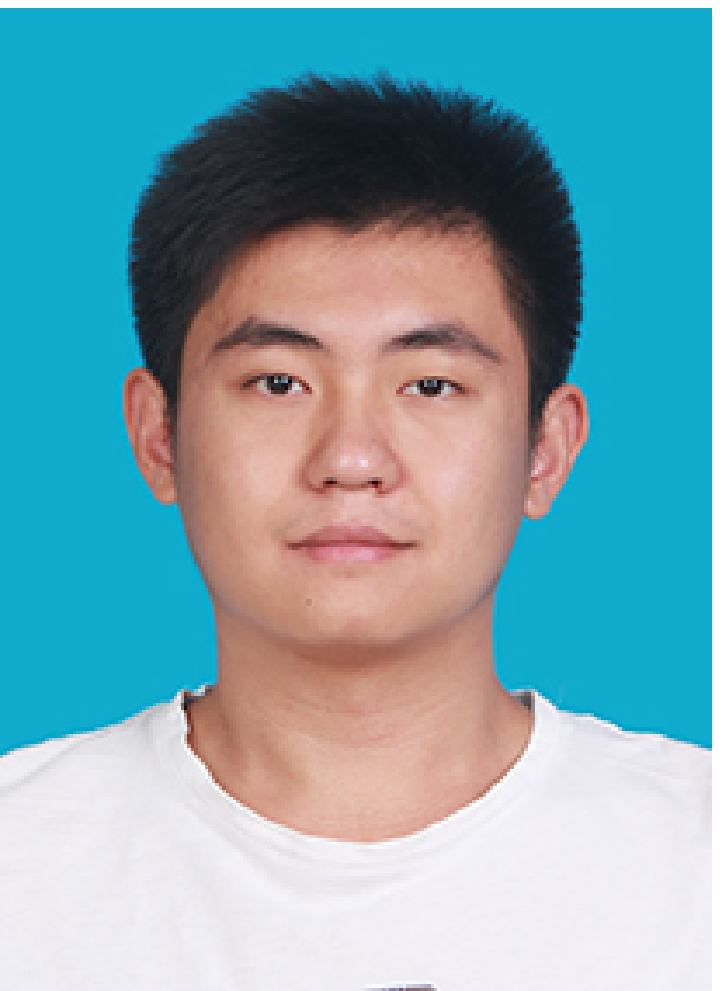}{Xinyu Gao}
received the B.E. degree of Communication	Engineering from Harbin Institute of Technology, Heilongjiang, China in 2014 and the PhD degree of Electronic Engineering from Tsinghua University, Beijing, China in 2019 (with the highest honor). He is currently working as a senior engineer for Huawei Technology, Beijing, China. His research interests include massive MIMO and mmWave communications, with the emphasis on signal processing. He has published more than 20 IEEE journal and conference papers, such as IEEE Journal on Selected Areas in Communications, IEEE Transaction on Signal Processing, IEEE ICC, IEEE GLOBECOM, etc. He has won the WCSP Best Paper Award and the IEEE ICC Best Paper Award in 2016 and 2018, respectively.
\end{CCJNLbiography}

\begin{CCJNLbiography}{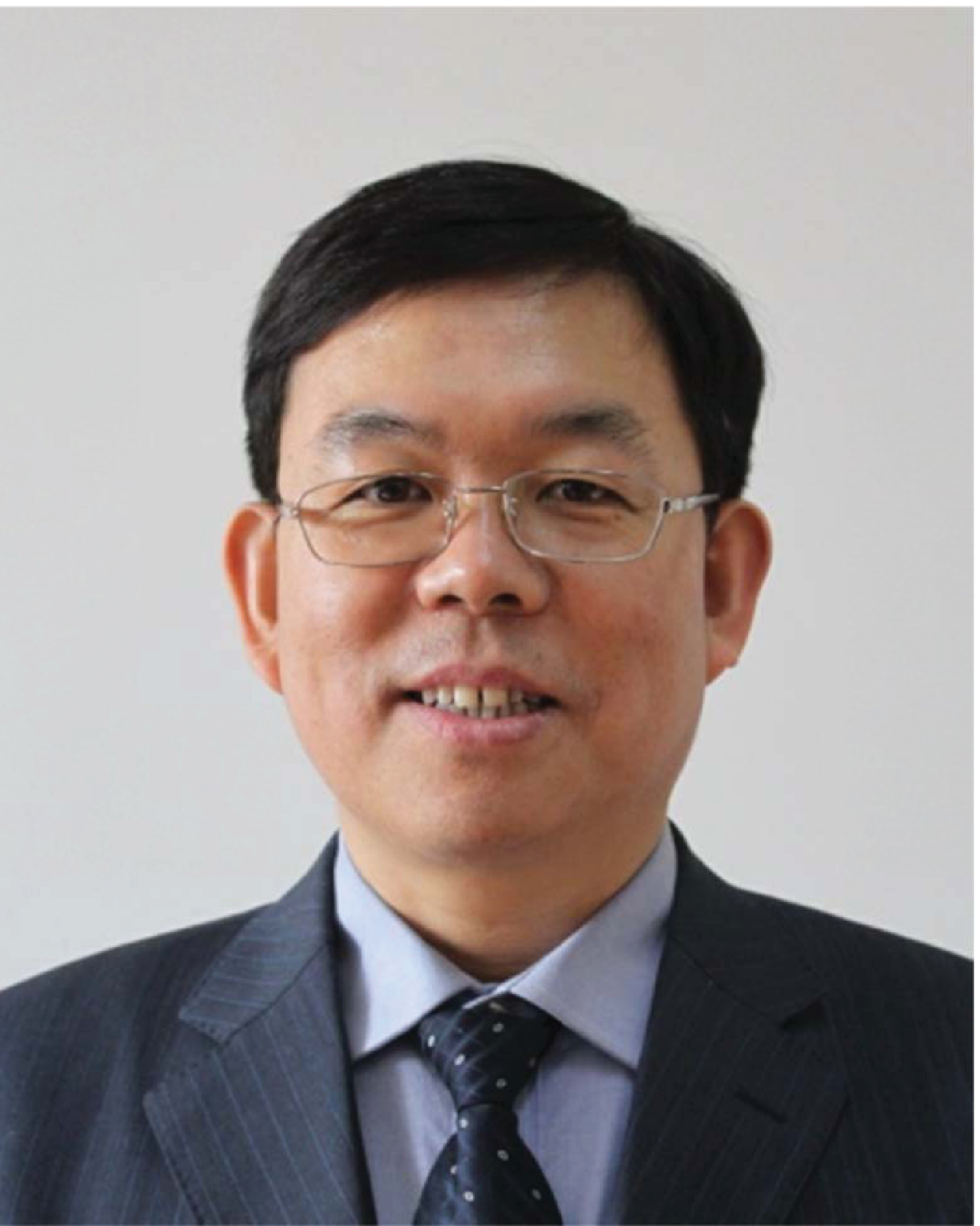}{Jianping An}
received the B.E. degree from Information Engineering University in 1987, and the M.S. and Ph.D. degrees from Beijing Institute of Technology, in 1992 and 1996, respectively. Since 1996, he has been with the School of Information and Electronics, Beijing Institute of Technology, where he now holds the post of Full Professor. From 2010 to 2011, he was a Visiting Professor at University of California, San Diego. He has published more than 150 journal and conference articles and holds (or co-holds) more than 50 patents. He has received various awards for his academic achievements and the resultant industrial influences, including the National Award for Scientific and Technological Progress of China (1997) and the Excellent Young Teacher Award by the China's Ministry of Education (2000). Since 2010, he has been serving as a Chief Reviewing Expert for the Information Technology Division, National Scientific Foundation of China. Prof. An's current research interest is focused on digital signal processing theory and algorithms for communication systems.
\end{CCJNLbiography}

\end{document}